\documentclass[12pt]{article}
\usepackage{graphicx}

\def\mpi{Max-Plank-Insitut f\"ur Physik\\
  F\"oringer Ring 6\\
  M\"unchen 80805 GERMANY}
\def\ec{Technische Universit\"at M\"unchen\\
  Excellence Cluster Universe\\
  Boltzmannstra{\ss}e 2\\
  Garching 85748 GERMANY}

\def\Title#1{\begin{center} {\Large #1 } \end{center}}
\def\Author#1{\begin{center}{ \sc #1} \end{center}}
\def\Address#1{\begin{center}{ \it #1} \end{center}}

\newenvironment{Abstract}{\begin{quotation}  }{\end{quotation}}
\newenvironment{Presented}{\begin{quotation} \begin{center} 
             PRESENTED AT\end{center}\bigskip 
      \begin{center}\begin{large}}{\end{large}\end{center} \end{quotation}}

\def\beq{\begin{equation}}
\def\eeq#1{\label{#1}\end{equation}}
\def\eeqn{\end{equation}}

\def\beqa{\begin{eqnarray}}
\def\eeqa#1{\label{#1}\end{eqnarray}}
\def\eeqan{\end{eqnarray}}

\let\bar=\overbar

\def\Dslash{\not{\hbox{\kern-4pt $D$}}}
\def\dslash{\not{\hbox{\kern-2pt $\del$}}}

\def\msb{{\bar{\ssstyle M \kern -1pt S}}}

\newcommand{\pipi}{\ensuremath{B \rightarrow \pi \pi}}
\newcommand{\pippim}{\ensuremath{B^{0} \rightarrow \pi^{+} \pi^{-}}}

\newcommand{\rhopi}{\ensuremath{B^{0} \rightarrow (\rho \pi)^{0}}}
\newcommand{\rhorho}{\ensuremath{B \rightarrow \rho \rho}}
\newcommand{\rhoprhom}{\ensuremath{B^{0} \rightarrow \rho^{+} \rho^{-}}}
\newcommand{\rhozrhoz}{\ensuremath{B^{0} \rightarrow \rho^{0} \rho^{0}}}
\newcommand{\rhoprhoz}{\ensuremath{B^{+} \rightarrow \rho^{+} \rho^{0}}}

\newcommand{\Ep}{\ensuremath{e^{+}}}
\newcommand{\Em}{\ensuremath{e^{-}}}

\newcommand{\aone}{\ensuremath{a_{1}(1260)}}
\newcommand{\Bp}{\ensuremath{B^{+}}}
\newcommand{\Bz}{\ensuremath{B^{0}}}
\newcommand{\Bzb}{\ensuremath{\bar{B}^{0}}}
\newcommand{\YFS}{\ensuremath{\Upsilon(4S)}}

\newcommand{\BBbar}{\ensuremath{B\bar{B}}}

\newcommand{\Dt}{\ensuremath{\Delta t}}

\newcommand{\Acp}{\ensuremath{{A}_{CP}}}

\newcommand{\Scp}{\ensuremath{{S}_{CP}}}

\newcommand{\phitwo}{\ensuremath{\phi_{2}}}

\begin{document}
\begin{titlepage}

\vfill
\Title{Measurement of $\alpha$/$\phitwo$ in $\pipi$, $\rho\pi$ and $\rho\rho$}
\vfill
\Author{ J.~Dalseno}
\Address{\mpi}
\Address{\ec}
\vfill
\begin{Abstract}
We present a summary of the measurements of the CKM angle, $\alpha$ (\phitwo), performed by the BaBar and Belle experiments which collect \BBbar\ pairs at the \YFS\ resonance produced in asymmetric \Ep\Em\ collisions. We discuss the measurements of the branching fractions and $CP$ asymmetries in the $B \rightarrow \pi \pi$, $\rho \pi$ and $\rho \rho$ final states that lead to constraints on $\alpha$ (\phitwo).
\end{Abstract}
\vfill
\begin{Presented}
Proceedings of CKM2010, the 6th International Workshop on the CKM Unitarity Triangle\\
University of Warwick, UK, 6-10 September 2010
\end{Presented}
\vfill
\end{titlepage}
\def\thefootnote{\fnsymbol{footnote}}
\setcounter{footnote}{0}

\section{Introduction}
The main goal of the BaBar experiment at SLAC and the Belle experiment at KEK is to constrain the unitarity triangle for $B$ decays. This allows us to test the Cabibbo-Kobayashi-Maskawa (CKM) mechanism for violation of the combined charge-parity ($CP$) symmetry~\cite{C,KM}, as well as search for new physics effects beyond the Standard Model (SM). These proceedings give a summary of the experimental status of measurements of the CKM phase, $\alpha$, hitherto referred to as \phitwo, defined from CKM matrix elements as $\phitwo \equiv \arg(-V_{td}V^{*}_{tb})/(V_{ud}V^{*}_{ub})$.

First-order weak processes (tree) proceeding by $b \rightarrow u \bar{u} d$ quark transitions such as $\Bz \rightarrow \pi\pi$, $\rho\pi$, $\rho\rho$ and $\aone\pi$, are directly sensitive to \phitwo. In the quasi-two-body approach, CKM angles can be determined by measuring the time-dependent asymmetry between \Bz\ and \Bzb\ decays~\cite{CP}. For the decay sequence, $\YFS \rightarrow B_{CP}B_{\rm Tag} \rightarrow f_{CP}f_{\rm Tag}$, where one of the $B$ mesons decays at time, $t_{CP}$, to a $CP$ eigenstate, $f_{CP}$, and the other decays at time, $t_{\rm Tag}$, to a flavour specific final state, $f_{\rm Tag}$, with $q = +1 (-1)$ for $B_{\rm Tag} = \Bz (\Bzb)$, the decay rate has a time-dependence given by
\begin{equation}
  {P}(\Delta t, q) = \frac{e^{-|\Dt|/\tau_{B^0}}}{4\tau_{B^0}} \bigg[ 1+q(\Acp\cos\Delta m_d \Dt + \Scp\sin\Delta m_d \Dt) \bigg],
\end{equation}
where $\Dt \equiv t_{CP}- t_{\rm Tag}$ and $\Delta m_d$ is the mass difference between the $B_{H}$ and $B_{L}$ mass eigenstates. The parameters, \Acp\ and \Scp, describe direct and mixing-induced $CP$ violation, respectively. An alternate notation where $C_{CP} = -\Acp$ also exists in literature.

If a single first-order weak amplitude dominates the decay, then we expect $\Acp=0$ and $\Scp=\sin2\phitwo$. On the other hand, if second-order loop processes are present, then direct $CP$ violation is possible, $\Acp\neq 0$. Additionally, as these loop processes (penguins) are not directly proportional to $V_{ub}$, our measurement of \Scp\ does not directly determine \phitwo, rather, $\Scp=\sqrt{1-\Acp^{2}}\sin(2\phitwo-2\Delta \phitwo)$, where $\Delta \phitwo$ is the shift caused by the second order contributions.

Despite this, it is possible to determine $\Delta \phitwo$ in $\Bz \rightarrow h^{+} h^{-}$ with an $SU(2)$ isospin analysis by considering the set of three $B \rightarrow hh$ decays where $hh$ is either two pions or two longitudinally polarised $\rho s$~\cite{iso}. The $B \rightarrow hh$ amplitudes obey the complex triangle relations,
\begin{equation}
  A_{+0} = \frac{1}{\sqrt{2}}A_{+-} + A_{00}, \;\;\;\; \bar{A}_{-0} = \frac{1}{\sqrt{2}}\bar{A}_{+-} + \bar{A}_{00}.
  \label{eq_iso}
\end{equation}
Isospin arguments demonstrate that $\Bp \rightarrow h^{+} h^{0}$ is a pure first-order mode in the limit of neglecting electroweak penguins, thus these triangles share the same base, $A_{+0}=\bar{A}_{-0}$, and $\Delta \phitwo$ can be determined from the difference between the two triangles. This method has an inherent 8-fold discrete ambiguity in the determination of \phitwo.

\section{\pipi}
\begin{figure}[htb]
  \centering
  \includegraphics[width=.53\textwidth]{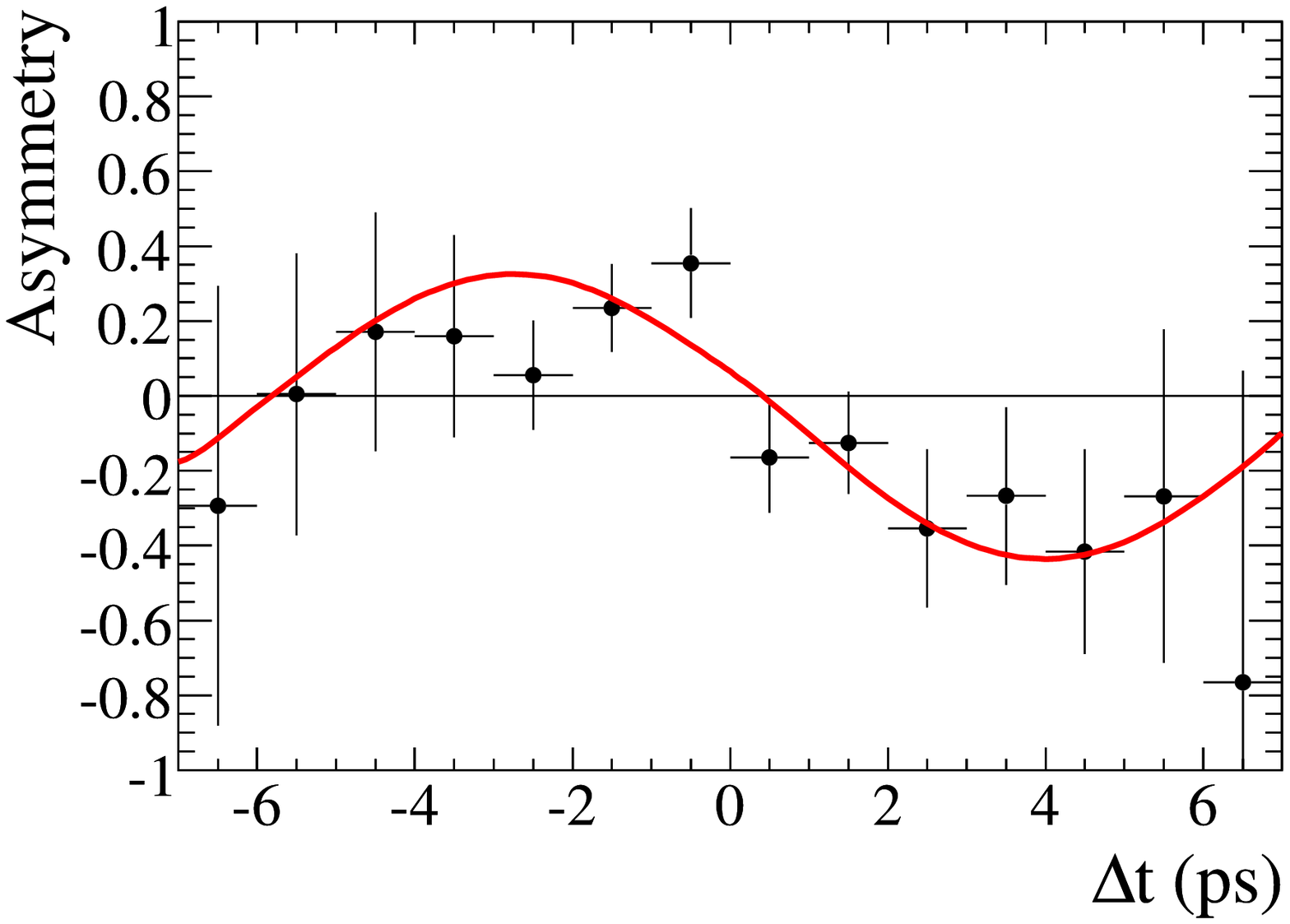}\hspace{20pt}
  \includegraphics[width=.38\textwidth]{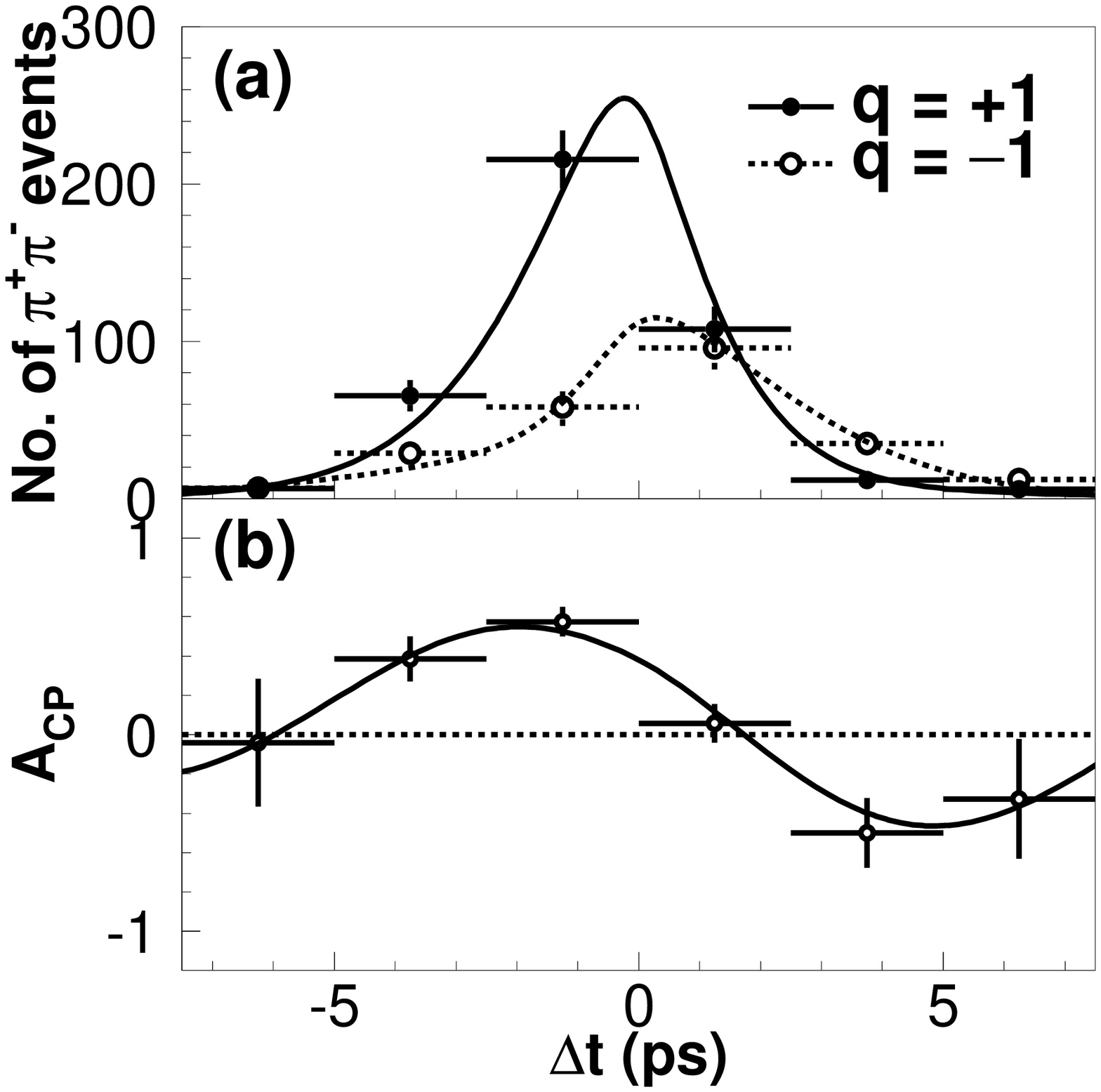}
  \caption{The left plot shows the time-dependent asymmetry, $a(\Dt) \equiv (N^{\rm Tag}_{\Bz} - N^{\rm Tag}_{\Bzb})/(N^{\rm Tag}_{\Bz} + N^{\rm Tag}_{\Bzb})$, of \pippim\ from BaBar. The right plots from Belle show the fit to \Dt\ for each flavour tag on top and the resulting asymmetry below. Mixing-induced $CP$ violation can be clearly seen in the asymmetry plots and the height difference in the \Dt\ projection indicates direct $CP$ violation.}
  \label{fig_pipi_tcpv}
\end{figure}

The analysis of \pipi\ performed by the BaBar collaboration is based on their full data set 467 million \BBbar\ pairs~\cite{pipi_BaBar}, while the analysis from the Belle collaboration is based on 535 million \BBbar\ pairs~\cite{pipi_Belle}. They obtain the $CP$ parameters,
$$\setlength\arraycolsep{0.2em}
\begin{array}{rclrcl}
  & & {\rm BaBar} & & & {\rm Belle}\\
  \Acp & = & +0.25 \pm 0.08 \pm 0.02 \; (3.0\sigma) & \hspace{30pt} \Acp & = & +0.55 \pm 0.08 \pm 0.05 \; (5.5\sigma)\\
  \Scp & = & -0.68 \pm 0.10 \pm 0.03 \; (6.3\sigma) & \hspace{30pt} \Scp & = & -0.61 \pm 0.10 \pm 0.04 \; (5.3\sigma),
\end{array}
$$
and the fit projections are shown in Fig.~\ref{fig_pipi_tcpv}. Both experiments have observed $CP$ violation in \pipi\ and the difference between the two measurements is 1.9$\sigma$. In the isospin analysis to remove the penguin contribution, BaBar excludes the range $[23^{\circ}, 67^{\circ}]$ at the 90\% CL and Belle excludes the range $[11^{\circ}, 79^{\circ}]$ at the 95\% CL.

\section{\rhorho}
\rhorho\ decays have an additional complication that the two spin-1 $\rho$ mesons have a relative orbital angular momentum, $L=0,1,2$. Since the $CP$ eigenvalue of \rhoprhom\ is $(-1)^{L}$, it is necessary to isolate a definite $CP$ component through angular analysis in order to constrain \phitwo. Conveniently, it turns out that the $\rho\rho$ system is dominated by the $CP$-even longitudinal amplitude~\cite{rhoprhom_BaBar, rhoprhom_Belle1}, which means the transverse component can be ignored in the constraint of \phitwo.

The BaBar analysis  of \rhoprhom\ is based on 384 million \BBbar\ pairs~\cite{rhoprhom_BaBar} while the Belle analysis is based on 535 million \BBbar\ pairs~\cite{rhoprhom_Belle2}. They obtain the $CP$ parameters,
$$\setlength\arraycolsep{0.2em}
\begin{array}{rclrcl}
  & & {\rm BaBar} & & & {\rm Belle}\\
  \Acp & = & 0.01 \pm 0.15 \pm 0.06 & \hspace{30pt} \Acp & = & +0.16 \pm 0.21 \pm 0.07\\
  \Scp & = & -0.17 \pm 0.20 ^{+0.05}_{-0.06} & \hspace{30pt} \Scp & = & +0.19 \pm 0.30 \pm 0.07.
\end{array}
$$

The BaBar collaboration has recently updated their \rhoprhoz\ analysis with their final data set~\cite{rhoprho0_BaBar}. They obtain the branching fraction, ${B}(\rhoprhoz) = (23.7 \pm 1.4 \pm 1.4) \times 10^{-6}$, which allows a precise measurement of the isospin triangle base, and $\Acp = 0.054 \pm 0.055 \pm ±0.010$, showing no evidence for amplitudes that do not conserve isospin.

Both collaboration have also presented result on \rhozrhoz. This is mode is experimentally difficult to isolate due to its relatively low branching fraction in the presence of multiple backgrounds with the same final state. BaBar has observed this mode with a significance of $3.1 \sigma$~\cite{rho0rho0_BaBar} and obtained the $CP$ parameters, $\Acp = -0.2 \pm 0.8 \pm 0.3$ and $\Scp = +0.3 \pm 0.7 \pm 0.2$, while Belle has obtained an upper limit~\cite{rho0rho0_Belle}.

\begin{figure}[htb]
  \centering
  \includegraphics[width=.35\textwidth]{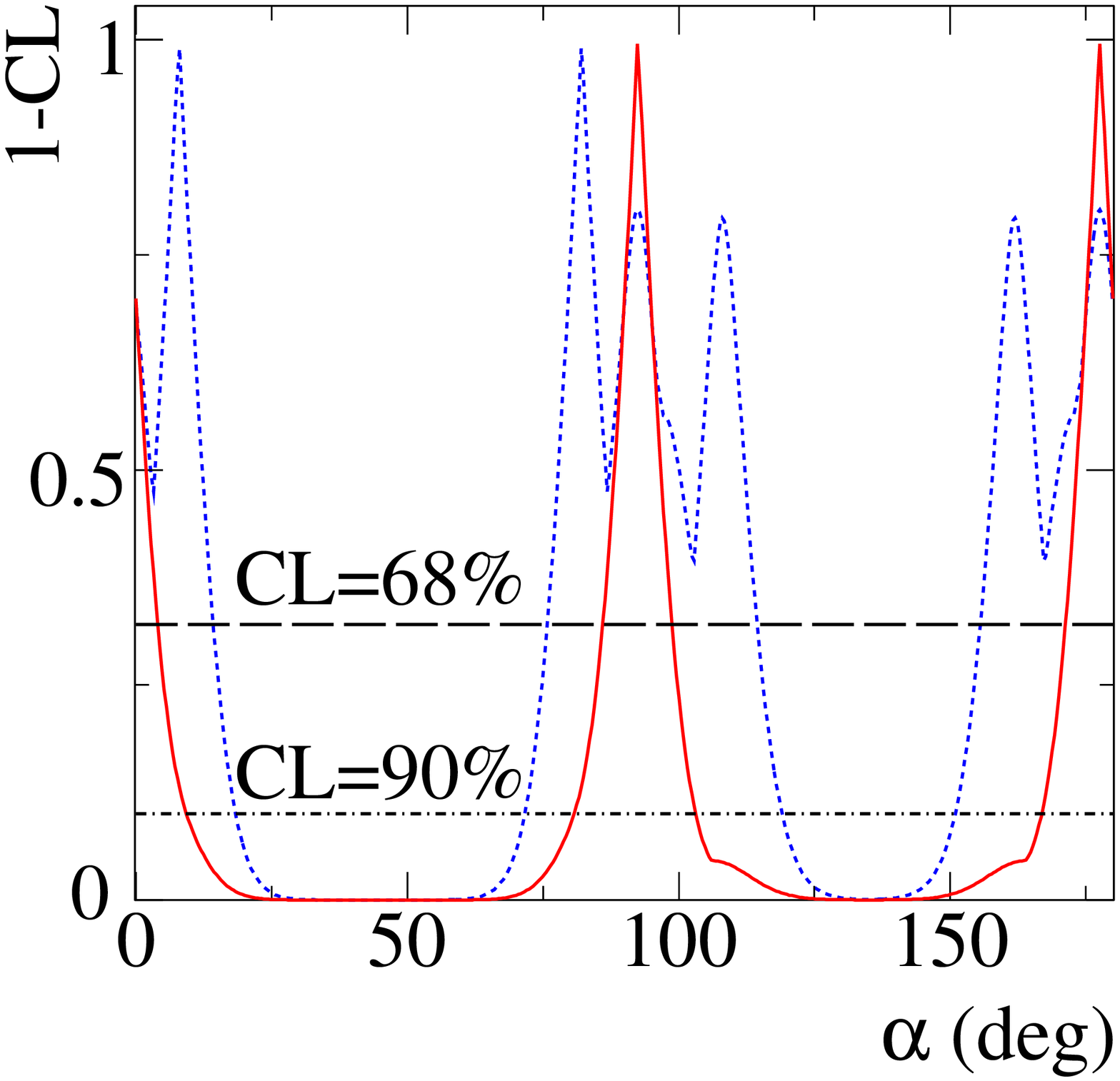}\hspace{10pt}
  \includegraphics[width=.6\textwidth,height=.34\textwidth]{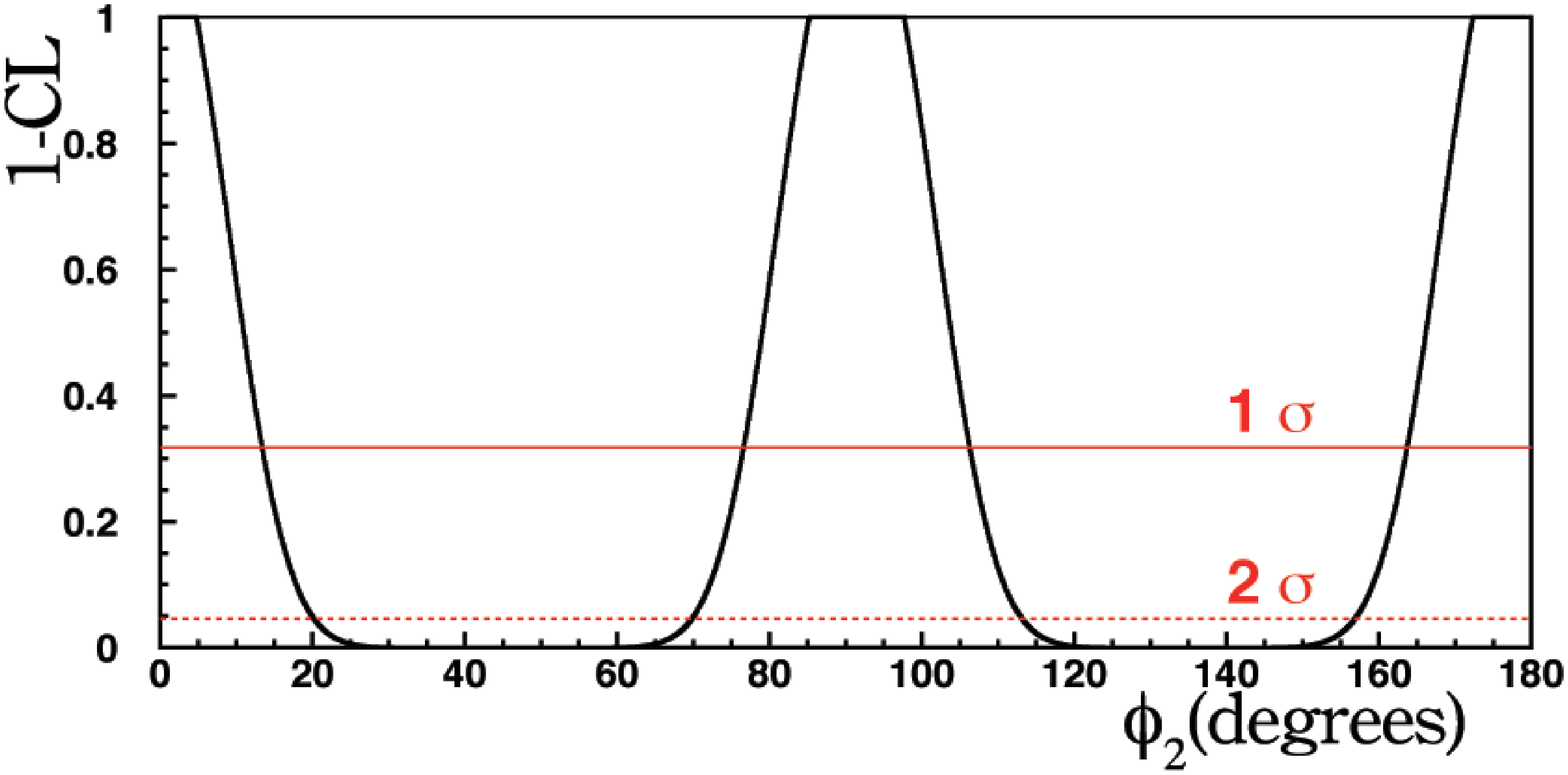}
  \caption{The left plot shows the constraint on \phitwo\ in the \rhorho\ system using only BaBar results. The right plot shows the constraint from Belle which uses its latest \rhozrhoz\ result, otherwise world averages. This analysis was performed before the recent update of \rhoprhoz\ from BaBar and a plateau is present as there is no constraint on $\Acp(\rho^{0}\rho^{0})$.}
  \label{fig_rhorho_phi2}
\end{figure}

A consequence of the small \rhozrhoz\ branching fraction relative to \rhoprhoz, is that the isospin triangles become flat making the 4 solutions of  $\Delta \phitwo$ nearly degenerate. The constraints on \phitwo\ in \rhorho\ are shown in Fig.~\ref{fig_rhorho_phi2} from which BaBar determines $\phitwo = (92.4 ^{+6.0}_{-6.5})^{\circ}$ and Belle finds $\phitwo = (91.7 \pm 14.9)^{\circ}$.

\section{\rhopi}
As \rhopi\ is not a $CP$ eigenstate, four flavour-charge configurations need to be considered. Unlike the previous channels, it is possible to constrain \phitwo\ explicitly in a time-dependent amplitude analysis that includes variations of the strong phase of interfering $\rho$ resonances over the Dalitz Plot~\cite{tdpa_phi2}. The relative moduli and phases of the six possible amplitudes of $\Bz (\Bzb) \rightarrow \pi^{+} \pi^{-} \pi^{0}$ decays via charged and neutral intermediate $\rho$ resonances are determined. These amplitudes are constructed from isospin relations from which \phitwo\ can be constrained without ambiguity.

\begin{figure}
  \centering
  \includegraphics[width=.43\textwidth]{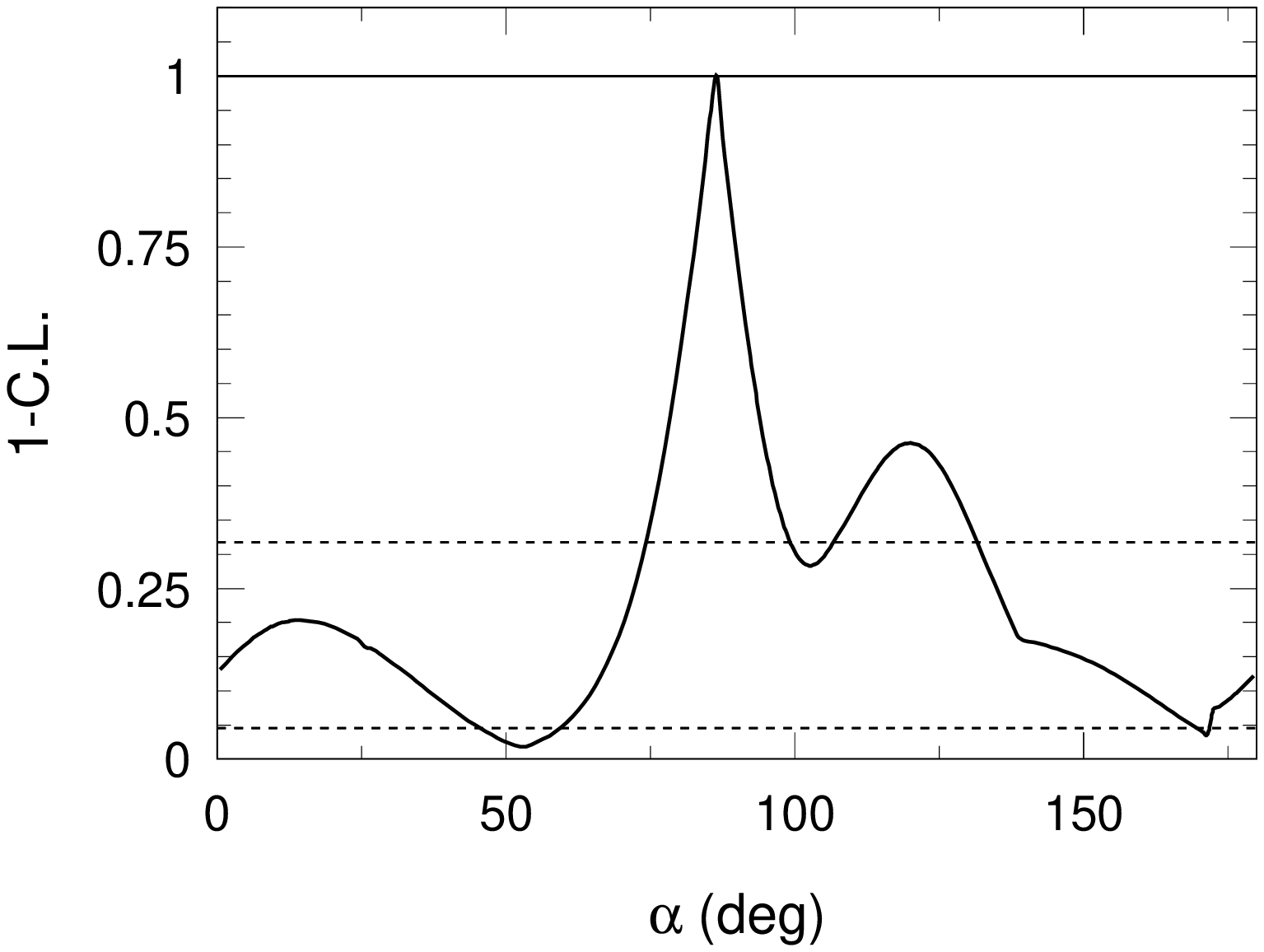}\hspace{20pt}
  \includegraphics[width=.45\textwidth,height=.34\textwidth]{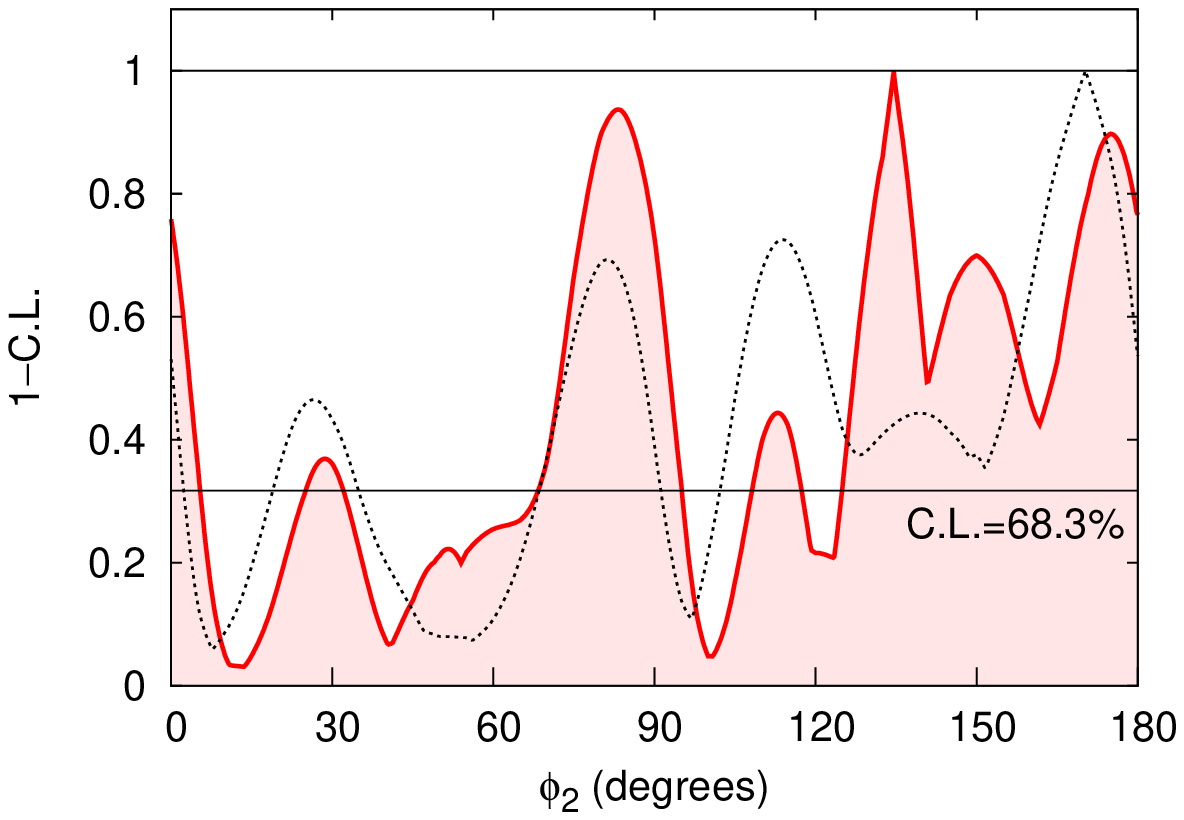}
  \caption{The left plot from BaBar shows the constraint on \phitwo\ in the \rhopi\ system with their time-dependent amplitude analysis. The right plot shows the constraint from Belle where the dashed curve corresponds to the BaBar curve and the red curve contains additional constraints from charged $\Bp \rightarrow (\rho \pi)^{+}$ modes.}
  \label{fig_rhopi}
\end{figure}

The BaBar and Belle collaborations have performed this analysis with 375 and 449 million \BBbar\ pairs, respectively, and are in good agreement. Their corresponding \phitwo\ scans are shown in Fig.~\ref{fig_rhopi} where BaBar obtains $\phitwo = (87^{+45}_{-13})^{\circ}$~\cite{rhopi_BaBar} while Belle can only constrain $68^{\circ} < \phitwo < 95^{\circ}$ at 68.3\% CL for the solution consistent with SM~\cite{rhopi_Belle}.

\section{Summary}
We have discussed the measurements of the branching fractions and $CP$ asymmetries in the $B \rightarrow \pi \pi$, $\rho \pi$ and $\rho \rho$ final states that lead to constraints on $\alpha$ (\phitwo). The current world averages of \phitwo\ as computed by the CKMfitter and UTfit collaborations are $\phitwo = (89.0^{+4.4}_{-4.2})^{\circ}$~\cite{CKMfitter} and $\phitwo = (91.4 \pm 6.1 )^{\circ}$~\cite{UTfit}, respectively. The next generation of $B$ physics experiments, BelleII and SuperB, have now received initial funding and are expected to reduce the uncertaity on \phitwo\ down to $1^{\circ}-2^{\circ}$~\cite{Belle2, SuperB}.

\end{document}